\documentstyle[floats,twocolumn,aps,prl] {revtex}
\begin{document}
\draft

\title{High-dimensional quantum dynamics
   of adsorption and desorption of H$_2$ at Cu(111)}

\author{Axel Gross, Bj{\o}rk Hammer\cite{Hamadr},
and Matthias Scheffler}

\address{Fritz-Haber-Institut der Max-Planck-Gesellschaft, Faradayweg 4-6,
D-14195 Berlin-Dahlem, Germany}

\author{Wilhelm Brenig}

\address{Physik-Department T30, Technische Universit\"at M\"unchen,
D-85747 Garching, Germany}

\maketitle

\begin{abstract}
We performed high-dimensional quantum dynamical calculations of the
dissociative adsorption and associative desorption of H$_2$/Cu(111).
The potential energy surface (PES) is obtained from
density functional theory calculations. Two regimes of dynamics are found,
at low energies sticking is determined by the minimum energy barrier,
at high energies by the distribution of barrier heights.
Experimental results are well-reproduced qualitatively, but
some quantitative discrepancies are identified as well.
\end{abstract}


Dissociative adsorption is one of the crucial steps in heterogenous
catalysis. This has motivated, apart from its fundamental interest for
surface chemistry, a large number of studies on well defined
model systems where the dissociative
adsorption process can be investigated in detail. The system which has been
studied to the largest extent is H$_2$/Cu(111) (see, e.g., \cite{Ren94,Hol94}).
For this system the adsorption is hindered by a noticable energy barrier.
It has been found that the dissociative adsorption of hydrogen approximately
obeys normal-energy scaling \cite{Ret92}, which means that the sticking
probability is only determined by the component of the molecule's momentum
perpendicular to the surface. Usually normal-energy scaling is associated
with a flat, structureless surface. For the (111) metal surface this
seemed to be an appropriate description as this is
the most densely packed surface.

Indeed the {\em electron density} at clean metal (111) surfaces
shows only a small corrugation of less than 0.1~{\AA} at a
distance of 1~{\AA} above the center of the surface layer \cite{Che75}.
However, adsorption properties are determined by the local chemical
interaction on the surface. In recent total energy calculations
\cite{Ham94} using density functional theory (DFT) together with the
generalized gradient approximation (GGA) \cite{Per92} for the
exchange-correlation functional, the potential energy
surface (PES) for H$_2$/Cu(111) has been determined. Besides the GGA the
main approximation is due to the super-cell approach and the {\bf k}-summation.
A rather strong corrugation is found  when all six degrees
of freedom of the H$_2$ molecule are varied.
For a 4-layer substrate with the molecular axis kept parallel to the surface,
which is the most favorable configuration for dissociation,
the energy barrier to adsorption varies between 0.73~eV and 1.43~eV
within the surface unit cell. Also the distance of the barrier
from the surface and the extension of the hydrogen bond at the barrier
region strongly depend on the point of impact in the unit cell and the
orientation of the molecule.

In order to investigate the consequences of the strong corrugation on the
adsorption and desorption dynamics, we have taken the GGA-PES as an input
for a quantum dynamical simulation. We use a five-dimensional parametrization
of the results obtained for the 4-layer substrate with the molecular axis being
kept parallel to the surface as for this configuration the lateral corrugation
of the PES has been determined in great detail in the DFT calculations.
Although the polar rotation of the molecule is not considered,
to our knowledge this is the first study of the influence of lateral
corrugation on the quantum dynamics of adsorption and desorption which
takes into account such a large number of
degrees of freedom using a realistic PES.
Because of the high-dimensionality of the calculations, the comparison of the
results with experimental data for the H$_2$/Cu(111) system will in addition
provide information about the accuracy of the theoretical GGA-PES.
We will show that the GGA-PES reproduces the experimental findings
qualitatively well, but some quantitative differences are identified. We will
demonstrate the importance of the corrugation of the PES for the
adsorption dynamics, the angular dependence of the sticking probability,
the mean kinetic energy in desorption, and the desorption flux.

The quantum dynamics is determined by solving the time-independent
Schr\"odinger equation for the H$_2$ molecule moving in the GGA-PES.
We use the concept of the {\em local reflection matrix} (LORE) \cite{Bre93}
and the {\em inverse local transmission matrix} (INTRA) \cite{Bre94}. The
propagation of the reflection and inverse transmission matrices on an accurate
piecewise constant representation of the PES avoids exponentially increasing
evanescent waves whereby numerical instabilities are prevented.
The center of mass distance of the molecule from the surface $Z$,
its two surface coordinates $X$ and $Y$ and the interatomic
molecular spacing $r$  are taken as dynamical variables.
The molecular azimuthal orientation is treated
as a fixed parameter in the adsorption simulation, and all theoretical
results are then obtained by averaging over twelve different
orientations. This is a reasonable
approximation because the molecular rotation is a slow process. For instance,
an H$_2$ molecules with a kinetic energy of 0.5~eV in the rotational
state $J = 4$ corresponding to an rotational
energy of 0.15~eV rotates only once in the time it takes to move forward
4.2~{\AA}, while the interaction range of the potential is 0.5~{\AA}.

The local nature of LORE and INTRA does not only give information about final
transition probabilities, but also about the scattering solution of the wave
function everywhere in space \cite{Bre94b}.
To give an idea about the size of the corrugation of our PES and its
consequences on the dynamics we have plotted in Fig.~\ref{wfplot}
the positive real part of the wave function for an
incident beam with total kinetic energy $E_i = 1.08$~eV and incident angle
$\theta_i = 45^{\circ}$ scattered at a surface with one-dimensional
lateral corrugation having the same variation
in energy and location of the barrier as the full potential. The
sticking probability is given by the flux of the particles that traverse the
barrier. To make both the incoming plus transmitted and the reflected part
of the beam visible in one {\em single} plot, we have restricted in the plot
the lateral extension of the beam perpendicular to its propagation direction,
which is in principle infinite. The splitting of the beam in a reflected and
a transmitted part can be traced very clearly. Also the development
of the different diffraction beams can be followed.

In Fig.~\ref{stick} we show results for the sticking probability as a function
of the normal kinetic energy of the incident H$_2$ beam initially in the
vibrational ground state for various polar angles of incidence
($\theta_i = 0^{\circ}$ refers to normal incidence).
Furthermore we present results for two-dimensional calculations corresponding
to
non-corrugated surfaces with the minimum and maximum barrier of the fully
corrugated surface. Also a sticking curve derived from experiment \cite{Mic91}
is shown. For $\theta_i = 0^{\circ}$ two energy regimes can be separated.
At low energies (Fig.~\ref{stick}b) the ``5D'' sticking curve
for the corrugated surface is parallel to the 2D sticking curve for the
non-corrugated surface with the minimum energy barrier. The 2D sticking curve
for the maximum barrier is suppressed by more than ten orders of magnitude
and has a different slope. These results indicate that the ``5D'' data can
be explained by a kind of key-hole effect. Since sticking at these energies
can only be achieved by tunneling, it is very sensitive to the barrier height.
The sticking for normal incidence at low energies is therefore
determined by the molecules which hit the surface within the lowest
energy configuration; molecules which hit the surface at unfavorable
sites or with unfavorable orientation
do practically not contribute to the sticking.

For incident energies larger than approximately 0.6~eV the slope of
the sticking curve for the corrugated surface in the logarithmic plot changes.
At this energy adsorption without tunneling becomes possible since the
effective minimum barrier height is reduced by 0.15~eV for molecules in their
vibrational ground state due to the softening of the H$_2$ bond at the surface
and the accompanied reduction of the zero-point vibrational energy.
In the linear plot (Fig.~\ref{stick}a) the sticking curve
obtained for the corrugated surface connects the onset of sticking for the
minimum barrier approximately linearly
with the point where the sticking probability in the maximum barrier system
reaches unity. Test calculations with a changed barrier widths but unchanged
barrier heights yield almost
unchanged sticking probabilities at these energies.
These facts indicate that the sticking probability in this energy range
is determined by the distribution of the barrier heights. This is the
first multidimensional evidence that for large energies the sticking
can be understood in terms of the region of the surface that classically is
available to dissociation, which is an important assumption in the so-called
hole-model \cite{Kar87}.

The comparison of the experimentally derived sticking curve with the
theoretical
results for the high-dimensional PES shows that the increase for sticking
probabilities between 0.01 and 0.4 is actually similiar. The main difference in
this range of sticking probabilities is that the theoretical results appear at
energies which are too high by about 0.2~eV. This indicates that the minimum
energy barrier for dissociation of the GGA-PES should be lowered from 0.73~eV
to
about 0.5~eV. In fact, convergence tests reported in Ref.~\cite{Ham94}
have shown that such a lowering of the barrier is to be expected.

For low and for high energies, however, the theoretical sticking probabilities
are at variance with the experiment.
The experimental curve has been derived by fitting existing
adsorption data and desorption data via the
principle of detailed balance to an assumed functional form of the
sticking curve \cite{Mic91}. In detailed balance the desorption flux is
determined by the product of the sticking probability, which increases
exponentially in the low-energy regime, with the exponentially decreasing
Boltzmann factor, resulting in a desorption flux which is strongly peaked at
energies close to the minimum barrier height.
Beam adsorption data, which are only available up to energies of about
0.5~eV for H$_2$ \cite{Mic91}, are dominated at low energies by the
initially vibrationally excited molecules \cite{Ret92}. Thus at least for
initially non-vibrating molecules the fitting procedure might not be very
sensitive at low and high energies.
For energies close to the minimum barrier height the agreement between
our results and experiment (except for the energy shift) suggests that the
spatial corrugation of our PES resembles that of reality. This also implies
that the width of the experimentally derived sticking curve does
not correspond to the barrier width \cite{Dar92}, but to the distribution
of barrier heights since it is this distribution which determines the increase
of the sticking probability in this range.

Just recently it has been pointed out \cite{Dar94b} that, apart from the change
in momentum of inertia, an isomorphism exists between surface corrugation and a
planar rotor. This can be used to give an estimate about the effect of
dynamically including the rotations. The total energy for a rotated H$_2$ dimer
has been calculated: At the optimum reaction configuration the energy
required to turn the molecule upright is 0.4~eV \cite{Ham94} which
is of the order of the lateral corrugation. Thus
taking into account the rotation will
have a similiar effect as including the surface corrugation, namely that
for $J = 0$ in the low energy regime the sticking probability will basically
be suppressed by a constant factor due to the smaller key-hole in phase space,
while in the high energy regime the energy at which the sticking probability
reaches one will be shifted to larger energies due to the existence of higher
barriers. The onset of the high energy regime will not be changed since the
minimum barrier will remain the same.

For non-normal incidence also two energy regimes can
be separated. At high energies (above 0.6~eV, see Fig.~\ref{stick}a)
the sticking obeys normal-energy scaling in agreement with the experiment
although the surface is strongly corrugated. This issue
has already been addressed by Darling and Holloway \cite{Dar94}.
Using model potentials they showed that the variation of the barrier height and
the variation of the barrier position have opposing
effects as far as the role of addional parallel momentum for adsorption is
concerned. If the maximum barriers are at a greater
distance from the surface these opposing effects cancel to a large extent
leading to approximate normal-energy scaling even for strongly corrugated
surfaces \cite{Dar94}. The GGA-PES shows these features
(see Fig.~\ref{wfplot}) indicating that this explanation also applies to our
calculations including the full lateral corrugation.

However, Fig.~\ref{stick}b reveals that in the low energy range,
which had not been considered in Ref.~\cite{Dar94},
additional parallel momentum helps to cross the barrier.
The situation is shown in Fig.~\ref{wfplot}. A large
part of the transmitted fraction of the beam propagates through the minimum
barrier site without moving like a classical particle, but being steered
in the tunneling process to the energetically most favorable propagation
path thereby converting parallel into normal energy, which in turn
increases the tunneling probability \cite{Gro94b}.

In Fig.~\ref{deskin} the mean kinetic energy of desorbing molecules
versus desorption angle at a surface temperature of T$_s$~=~925~K
is plotted. Measurements by Comsa and David \cite{Com82} have shown
that the kinetic energy is independent of
the desorption angle, which is surprising taking into account normal
energy scaling for H$_2$/Cu(111) in a classical model
(solid line in Fig.~\ref{deskin}).
This issue has been discussed at length by Michelsen and Auerbach
\cite{Mic91} who showed via detailed-balance arguments that due to the
significant width of the sticking probability, which effectively
changes with desorption angle, a constant mean kinetic energy in desorption
can be achieved even under the assumption of normal-energy scaling.
Our calculations also give a kinetic energy of desorbing molecules almost
independent of the desorption angle. Here it is mainly due to the
inclusion of geometric corrugation which makes parallel momentum more
effective for traversing the barrier leading to a more isotropic
distribution of desorbing molecules \cite{Gro94b}.

The mean kinetic energy obtained in our ``5D'' quantum dynamical calculation
is 0.2~eV larger than the one measured in experiments \cite{Com82,Ret93}.
This again corresponds to the shift in the minimum energy barrier height
of our PES as discussed above. The vibrational excitation of desorbing
molecules at T$_s$~=~925~K is calculated to be 0.87\% while the experimental
result is 2.9\% \cite{Ret93}. The GGA places the energy barrier for
dissociation correctly in the ``exit'' channel in the $rZ$-plane
(see Fig.~1a in \cite{Ham94}). However, the H$_2$ bond length
at the barrier top is only extended by 33\% so that the barrier position
is still too early \cite{Kue91,Gro94}.

Calculations using only a two-dimensional PES with the minimum energy path
yield 0.67~eV for the mean kinetic energy and
0.83~\% for the vibrational excitation in desorption at T$_s$~=~925~K.
These numbers are relatively close to the
ones derived for a fully corrugated surface indicating that most desorbing
molecules follow the minimum energy path. This shows that it is
justified to {\em neglect corrugation} in the simulation of {\em desorption
processes}  if there is one preferred reaction path to desorption which
is energetically much more favorable than all others.

The calculated angular distribution of desorbing molecules
can be fitted to a $\cos^n \theta$-curve with
$n \approx 25$. This distribution is narrower than the one seen in
experiments which yield an exponent of $n_{exp} \approx 8$ \cite{Com82}.
This can partly be attributed to the fact that the kinetic energy of desorbing
molecules is too high in our calculations. Furthermore, experiments have shown
that the angular distribution in desorption is broadened if the motion of the
surface atoms is taken into account \cite{Mic92}, which is neglected in our
calculations.

In conclusion, we have presented a dynamical study of the dissociative
adsorption and associative desorption in the system H$_2$/Cu(111)
using a five-dimensional potential energy surface which
has been determined by {\it ab initio} calculations in the generalized
gradient approximation.
Except for an energy shift of 0.2~eV due to the restricted super cell and
{\bf k}-point set, the GGA-PES describes the surface corrugation qualitatively
well. In the tunneling regime sticking is determined by the minimum energy
barrier. For energies larger than the effective minimum
barrier height the width of the sticking curve corresponds to the distribution
of barrier heights, and sticking obeys normal-energy scaling although
the PES is strongly corrugated. The mean kinetic energy of desorbing
molecules is almost independent of the desorption angle due to the
variation of the barrier location. In order to simulate
desorption experiments corrugation can be neglected if there is one
energetically favorable reaction path to desorption.

\begin{figure}[h]
   \caption{Positive real part of the wave function of H$_2$ scattered
            at a PES with one-dimensional lateral corrugation simulating
            H$_2$/Cu(111). The incident beam has a kinetic energy of
           $E_i = 1.08$~eV, the incident angle is $\theta_i = 45^{\circ}$.
          The contour spacing for the potential (thick lines) is 0.1~eV.
          The positions of the Cu atoms are at the potential maxima. }
\label{wfplot}
\end{figure}

\begin{figure}[h]
   \caption{Sticking probability versus normal kinetic energy for
molecules initially in the vibrational ground state.
a) Linear plot, b) logarithmic plot (note the different energy range).
5D-calculations for different incident angles at the corrugated
surface: solid line $\theta_i = 0^{\circ}$,
$\Diamond \ \theta_i = 15^{\circ}$, $\bigtriangleup \ \theta_i = 30^{\circ}$,
$\bigcirc \ \theta_i = 45^{\circ}$;
2D-calculations corresponding to a flat surface: dash-dotted line
for the minimum barrier, long-dashed line for the maximum barrier.
Dashed line: Experimental curve incorporating adsorption data and,
via the principle of detailed balance, also desorption data (from Ref.
\protect{\cite{Mic91}}). }
\label{stick}
\end{figure}

\begin{figure}[h]
   \caption{Mean kinetic energy of desorbing molecules versus desorption
    angle at a surface temperature of $T_s =925$~K obtained in our
    quantum dynamical calculation (open circles) and in a classical model
    for a flat surface (after Ref.~\protect{\cite{Com82}})
     with the minimum energy barrier of 0.73~eV.}
   \label{deskin}
\end{figure}


\begin{thebibliography}{99}
\bibitem[*]{Hamadr} Present address: Physics
Department, Technical University of Denmark, DK-2800 Lyngby, Denmark.
\bibitem{Ren94} K.D. Rendulic and A. Winkler, Surf. Sci. {\bf 299/300}, 261
(1994).
\bibitem{Hol94} S. Holloway, Surf. Sci. {\bf 299/300}, 656 (1994).
\bibitem{Ret92} C.T. Rettner, D.J. Auerbach, and H.A. Michelsen, Phys. Rev.
Lett. {\bf 68}, 1164 (1992).
\bibitem{Che75} J.R. Chelikowsky, M. Schl{\"u}ter, S.G. Louie, and M.L. Cohen,
Solid State Commun. {\bf 17}, 1103 (1975).
\bibitem{Ham94} B. Hammer, M. Scheffler, K.W. Jacobsen, and J.K. N{\o}rskov,
Phys. Rev. Lett., in press.
\bibitem{Per92}J.~P.~Perdew {\it et al.},
Phys.~Rev.~B {\bf 46}, 6671 (1992).
\bibitem{Bre93} W. Brenig, T. Brunner, A. Gross, and R. Russ, Z.~Phys.~B
{\bf 93}, 91 (1993).
\bibitem{Bre94} W. Brenig and R. Russ, Surf. Sci., in press.
\bibitem{Bre94b} W. Brenig, A. Gross, and R. Russ, to be published.
\bibitem{Mic91} H.A. Michelsen and D.J. Auerbach, J. Chem. Phys. {\bf 94},
 7502 (1991).
\bibitem{Kar87} M. Karikorpi, S. Holloway, N. Henriksen and J.K. N{\o}rskov,
Surf. Sci. {\bf 179}, L41 (1987).
\bibitem{Dar92} G.R. Darling and S. Holloway, J. Chem. Phys. {\bf 97}, 734
(1992).
\bibitem{Dar94b} G.R. Darling and S. Holloway, J. Chem. Phys. {\bf 101}, 3268
(1994).
\bibitem{Dar94} G.R. Darling and S. Holloway, Surf. Sci. {\bf 304}, L461
(1994).
\bibitem{Gro94b} A. Gross, subm to J. Chem. Phys.
\bibitem{Com82} G. Comsa and R. David, Surf. Sci. {\bf 117}, 77 (1982).
\bibitem{Ret93} C.T. Rettner, H.A. Michelsen, and D.J. Auerbach, J. Vac. Sci.
Technol. A 11, 1901 (1993).
\bibitem{Kue91} S. K\"uchenhoff, W. Brenig, and Y. Chiba, Surf. Sci. {\bf 245},
389 (1991).
\bibitem{Gro94} A. Gross, Surf. Sci. {\bf 314}, L843 (1994).
\bibitem{Mic92} H.A. Michelsen, C.T. Rettner, and D.J. Auerbach, Surf. Sci.
{\bf 272}, 65 (1992).



\end{thebibliography}
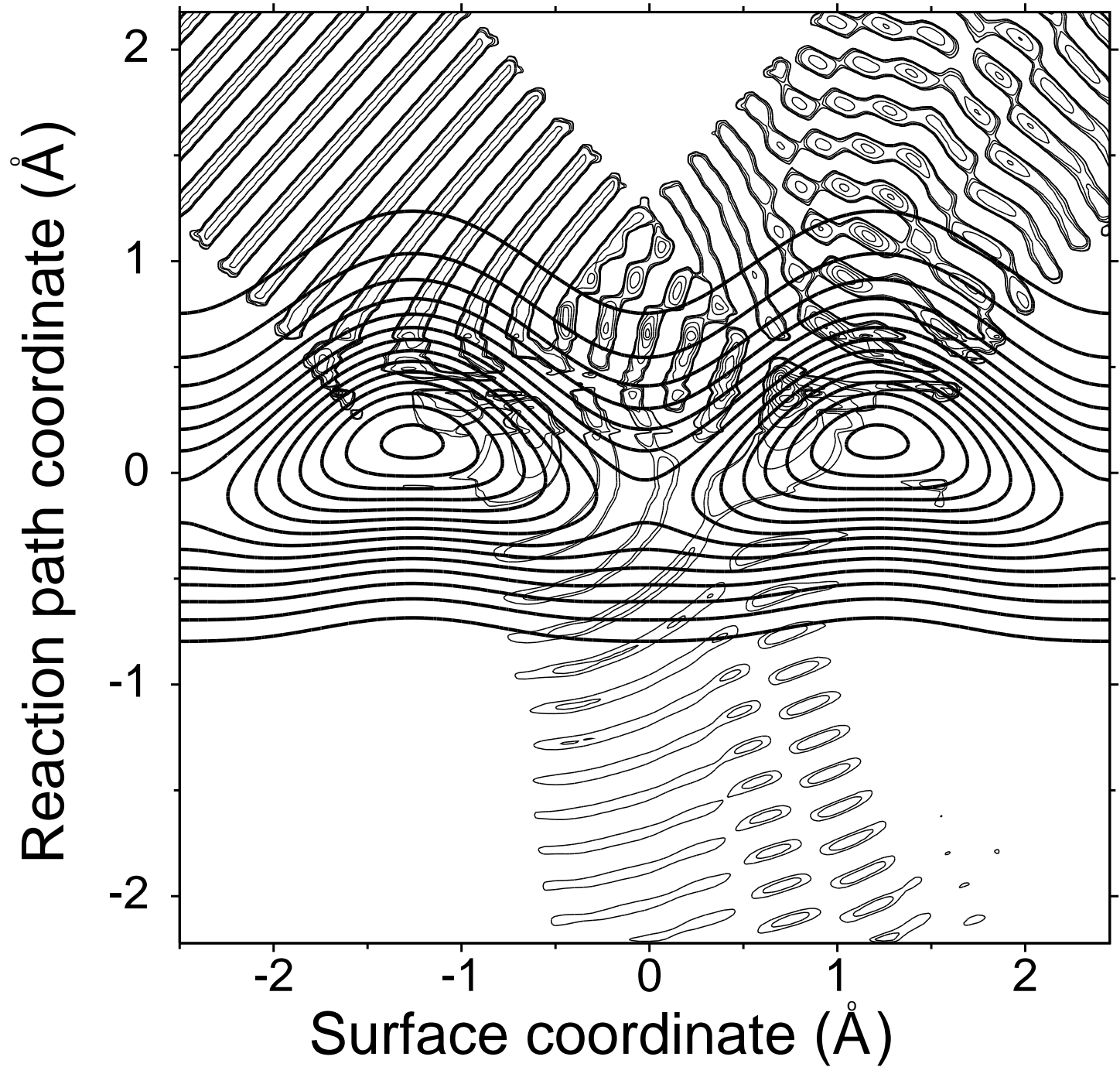
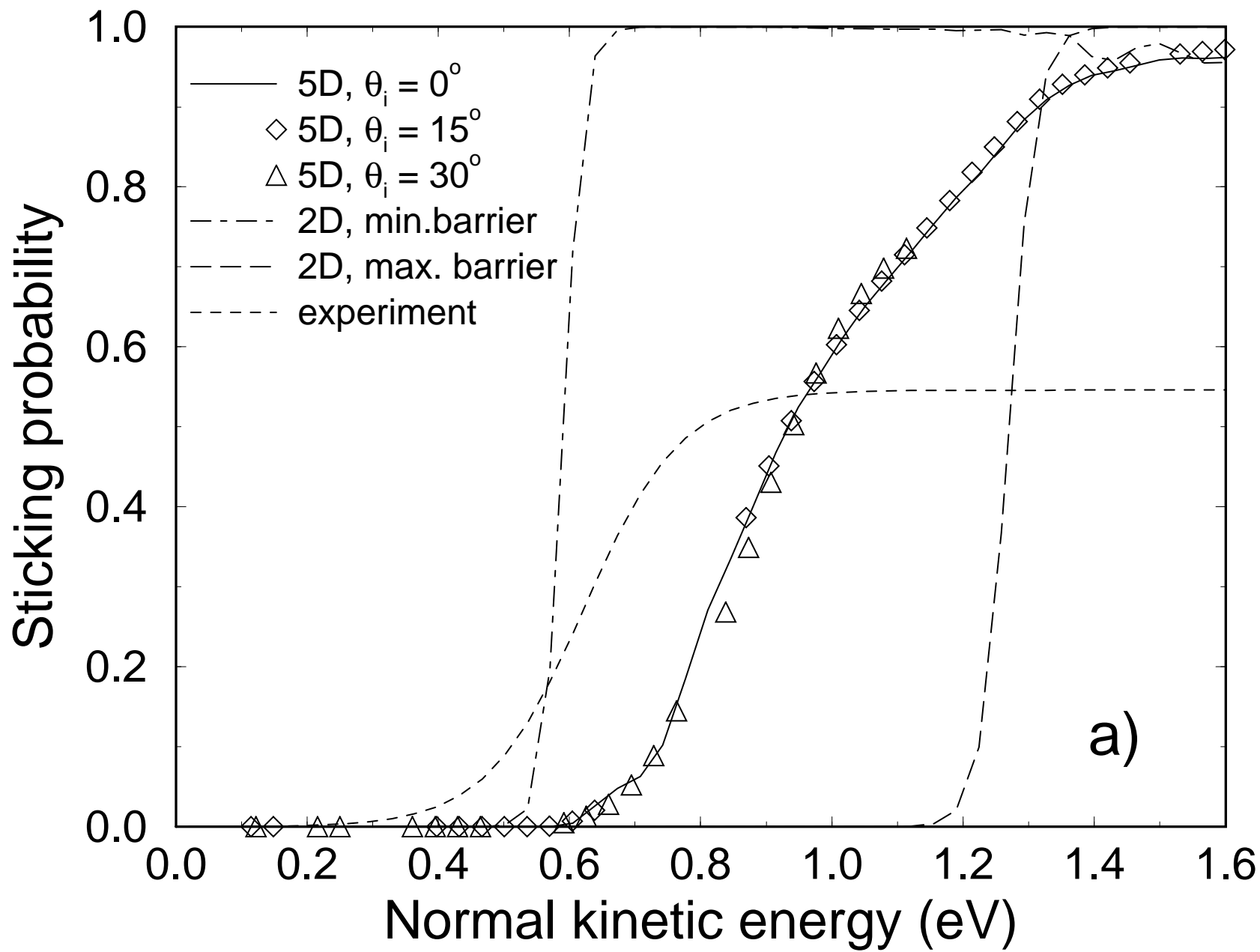
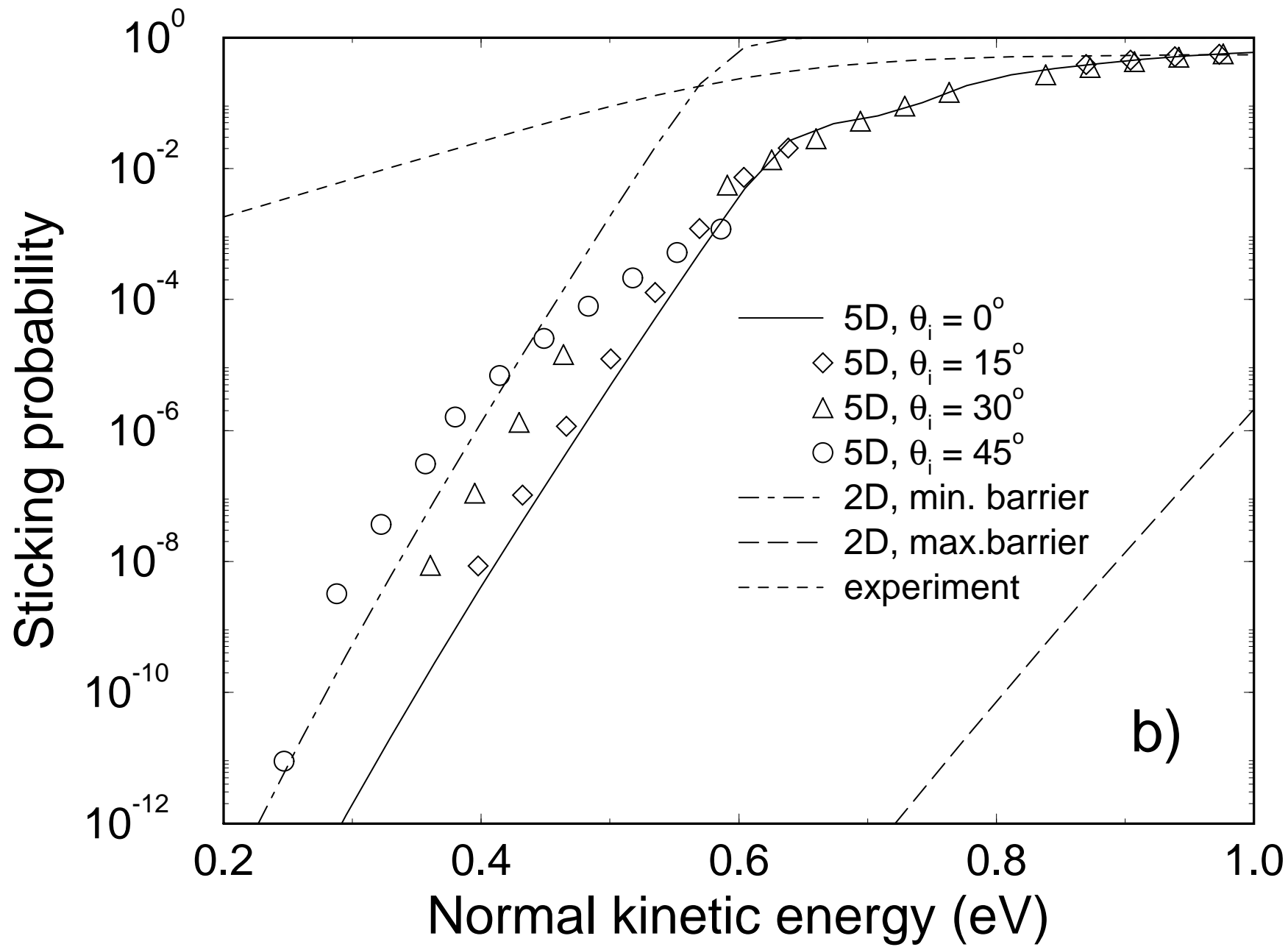
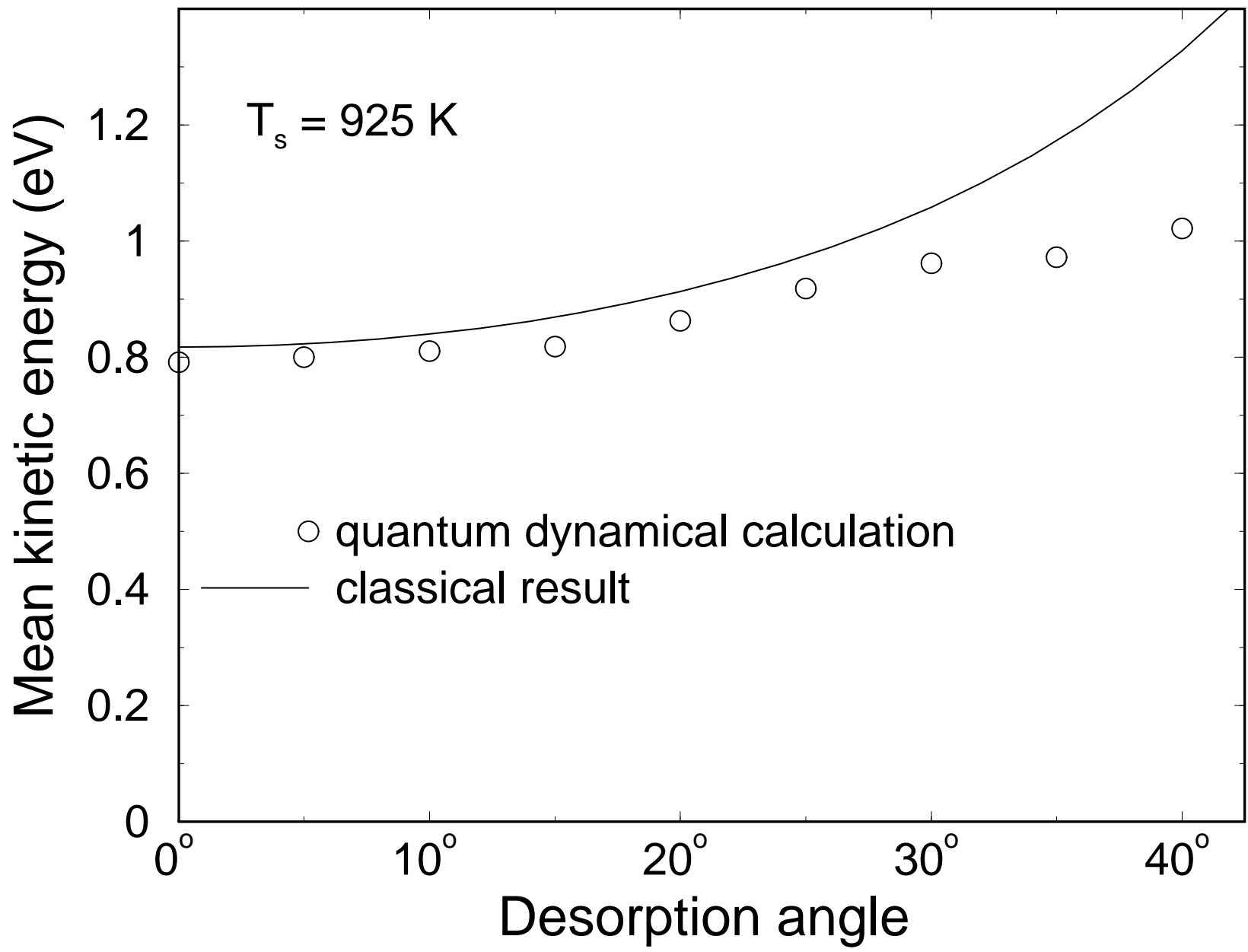
\end{document}